\documentclass[twocolumn,preprintnumbers,amsmath,amssymb]{revtex4}

\usepackage{graphicx}
\usepackage{dcolumn}
\usepackage{bm}
\usepackage{epstopdf}

\begin{document}

\preprint{APS/123-QED}

\title{The weak core and the structure of elites in social multiplex networks}

\author{Bernat Corominas-Murtra$^1$ and Stefan Thurner$^{1,2,3}$}

\thanks{Author correspondence: stefan.thurner@meduniwien.ac.at}

\affiliation{
$^1$ Section for Science of Complex Systems; Medical University of Vienna, Spitalgasse 23; A-1090, Austria\\
$^2$Santa Fe Institute, 1399 Hyde Park Road, New Mexico 87501, USA\\
$^3$ IIASA, Schlossplatz  1, A-2361 Laxenburg; Austria}

\begin{abstract}
Recent approaches on elite identification highlighted the important role of {\em intermediaries}, by means of a new definition of the core of a multiplex network, the {\em generalised} $K$-core. This newly introduced core subgraph crucially incorporates those individuals who, in spite of not being very connected, maintain the cohesiveness and plasticity of the core. Interestingly, it has been shown that the performance on elite identification of the generalised $K$-core is sensibly better that the standard $K$-core. Here we go further: Over a multiplex social system, we isolate the community structure of the generalised $K$-core and we identify the weakly connected regions acting as bridges between core communities, ensuring the cohesiveness and connectivity of the core region. This gluing region is the {\em Weak core} of the multiplex system. We test the suitability of our method on data from the society of 420.000 players of the Massive Multiplayer Online Game {\em Pardus}. Results show that the generalised $K$-core displays a clearly identifiable community structure and that  the weak core gluing the core communities shows very low connectivity and clustering. Nonetheless, despite its low connectivity, the weak core forms a unique, cohesive structure. In addition, we find that members populating the weak core have the best scores on social performance, when compared to the other elements of the generalised $K$-core. 
The weak core provides a new angle on understanding the social structure of elites, highlighting those subgroups of individuals whose role is to glue different communities
in the core. 
\end{abstract}

\maketitle

\section{Introduction}

Which social network structures within a social system define an elite? Elites are typically formed from individuals that have the capacity to accumulate large amounts of wealth,
power and influence. The location within the social multiplex network of social interactions enables this small group of people to have significant influence and control 
over a large fraction of the population. A crucial feature of elites is that relations between its members define a highly cohesive network at different levels. 
Its defining traits are still under discussion \cite{Mills:1956, Mills:1958, Keller:1963, Domhoff:1967, Bottomore:1993}. Intuitively, elite structures are  
formed by individuals with a large number of ties connecting them to the overall society and by individuals who, in spite not being highly connected, link the highly connected ones. 
The later can be seen as  {\em intermediaries} \cite{Friedkin:1984, Corominas-Murtra:2014}. 
\begin{figure*}
\begin{center}
\includegraphics[width=18.0cm]{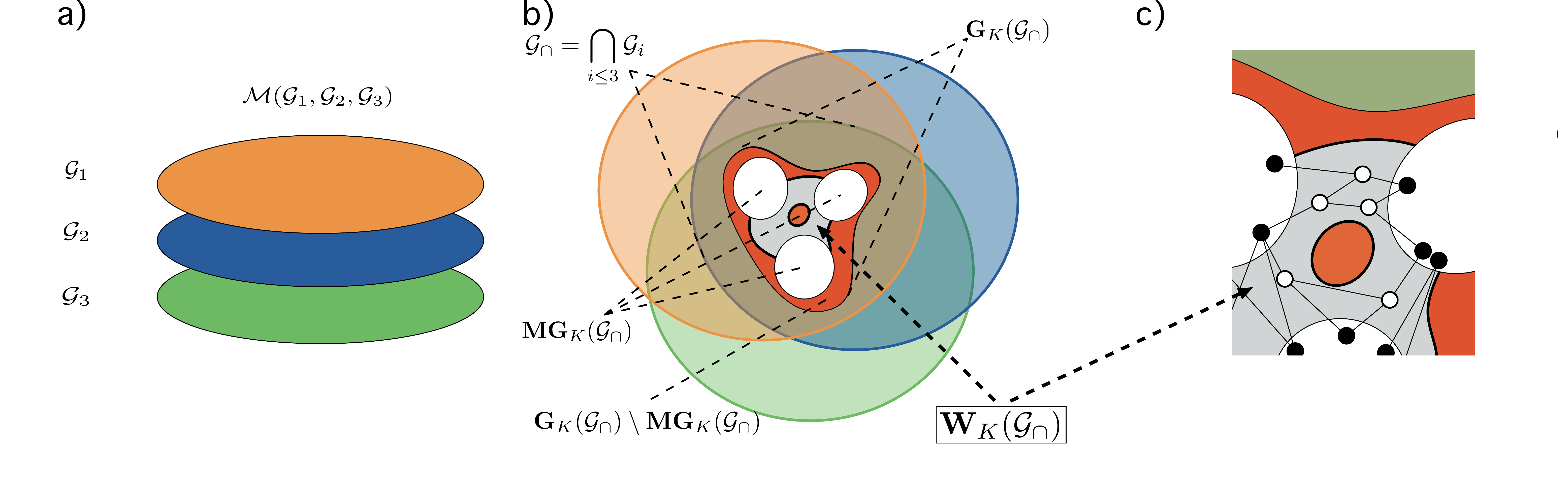}
\caption{A given multiplex system is composed by three layers, ${\cal G}_1, {\cal G}_2$ and ${\cal G}_3$ (a). Extracting the weak core 
(b): First, compute the intersection among the layers, ${\cal G}_{\cap}$, then compute the $G_K$-core of ${\cal G}_{\cap}$, namely $G_K({\cal G}_{\cap})$, 
depicted as the red region containing both grey and white components. After that we extract the $\mathbf{M}$-core of the $\mathbf{G}_K$-core 
($M=2$) thereby obtaining the subgraph $\mathbf{MG}_K({\cal G}_{\cap})$, whose components are  shown as white regions at the core. 
These three regions depict the communities defined through a high degree of clustering;  we call them the {\em core communities}. The {\em weak core} 
of ${\cal G}$, $\mathbf{W}_K({\cal G})$ (grey region), is the subgraph formed by all links and nodes that start in one of the core communities 
and end in another core community (grey region). No links between members of the same core community are 
allowed in $\mathbf{W}_K({\cal G})$. 
In (c) we show some examples of potential structures forming the $\mathbf{W}_K({\cal G})$-core. We differentiated the 
nodes belonging to $\mathbf{MG}_K({\cal G}_{\cap})$ (black) and to $\mathbf{G}_K({\cal G}_{\cap})\setminus\mathbf{MG}_K({\cal G}_{\cap})$ (white), 
to emphasise the hybrid, glue-like character of the weak core.}
\label{fig:Schema}
\end{center}
\end{figure*}

A social system can be fairly described  with a multiplex network (MPN) approach  \cite{Mucha:2010, Szell:2010b, Bianconi:2013a}. 
In a multiplex network, nodes interact through different types of relations or links. 
In this paradigm, elites have been thought to  form a 
cohesive region which organises the whole topology of the multiplex system \cite{Wasserman:1994}. A few decades ago, quantitative sociology 
developed the concept of the $K$-core to identify this small subset of highly influencial individuals \cite{Seidman:1983, Bollobas:1984, Dorogovtsev:2006}. 
Generally members of the $K$-core tend to be highly connected (hubs). The strong-connectivity requirement in the definition of the $K$-core, does not allow to identify  
the potentially important intermediaries or connectors. To improve this situation a {\em Generalised} $K$-core was suggested which includes connectors 
in the definition of the core of a complex network  \cite{Corominas-Murtra:2014}. The suitability of this definition was demonstrated in a virtual society of players of the 
Massive Multiplayer Online Game (MMOG) {\em Pardus}, and was compared to the classic $K$-core for the identification of elites. 
The incorporation of connectors provides a much richer description of the core. 

In this publication we want to take the next logical step and analyse the substructure of elites. In particular we will focus on the 
weakly connected regions of the core, which provide the `glue' for the different core communities. We expand the concept of a connector to an abstract 
structure which keeps the cohesiveness of the core of the multiplex network. The resulting subgraph we call the {\em weak core}, which defines the region 
of the core which prevents the core to disintegrate into its potential subcommunities. Interestingly, the notion of a weak core is independent of 
the definition of core and independent of the used community detection method.

We demonstrate our idea with data from the MMOG society of the game {\em Pardus} (http://www.pardus.at) \cite{Szell:2010a}, an open-ended online game 
with a worldwide player base which currently contains more than 420,000 people. MMOGs have been shown to be exceptional platforms over 
which quantitative results about social structures, dynamics, and organisational rules can be derived 
\cite{Corominas-Murtra:2014, Castronova:2005, Szell:2010a, Szell:2012a, Szell:2012b, Thurner:2012, Szell:2013, Fuchs:2014}. 
In this game players live in a virtual, futuristic universe where they interact with other players in a multitude of ways to achieve their 
self-posed goals.  A number of social networks can be extracted from the {\em Pardus} game, so that a dynamical multiplex network of a human social system can be 
quantitatively defined. The MPN consists of the time-varying communication, friendship, trading, enmity, attack, and revenge networks. 
Our findings in the virtual {\em Pardus} society confirm that indeed the weak core plays a crucial role in keeping the cohesiveness of the core of 
the multiplex system and, show that members populating this subgraph are characterised by the largest scores in quantitative 
social performance indicators. The weak core might be a crucial and practical step towards the understanding of the internal structures of elites.

The paper is organised as follows: In section \ref{sect2A} we formally define the multiplex network, in \ref{sect2B} 
we revisit the concepts of {\em generalised} $K$-core and the $M$-core, which will be used as a community structure detector. 
Section \ref{sect2C} introduces the concept of the weak core. In section \ref{sect2D} we discuss and define criteria to identify relevant 
levels of core organisation. Section \ref{sect3} presents the results for the weak core analysis in the {\em Pardus} society.  
In \ref{sect3A} we discuss topological aspects, and in \ref{sect3B} the social performance indicators of those individuals in the weak core 
are compared to those comprising other social groups. Finally, in section \ref{sect4} we discuss the results.

\section{Identification of  the Weak Core}
\label{sect2}
We introduce the following notation. We use  bold letters for the various core subgraphs, namely  
$\mathbf{K}$-core for the usual $K$-core subgraph, 
$\mathbf{G}_K$-core for the {\em generalised} $K$-core, 
$\mathbf{M}$-core for the $M$-core and 
$\mathbf{MG}_K$ for the $M$-core of a generalised $K$-core. In general, we will use the word {\em core} to refer to the $\mathbf{G}_K$-core.

A multiplex system ${\cal M}$ is made of $\mu$ layers, which represent different types of interactions or relations among the same set of nodes. 
Nodes are usually people; for the multiplex we write
\begin{equation}
{\cal M}={\cal M}({\cal G}_1,. . .,{\cal G}_\mu).
\end{equation}
Levels or layers of the multiplex are indexes by greek letters.
Figure \ref{fig:Schema} gives a schematic picture of the multiplex and the procedure described in the following. 

\subsection{Intersection of levels in a Multiplex system  \label{sect2A}}

Each layer of the multiplex  can be seen as a network ${\cal G}_{\alpha}(V, E_{\alpha})$ whose set of nodes $V$ is 
shared with the other layers ${\cal G}_{1}, . . .,{\cal G}_{\mu}$ and whose set of links $E_{\alpha}$ describes the particular 
connections that occur at level $\alpha$. The number of nodes of the multiplex system will be denoted by $|V|$ and the number of links of a given level $\alpha$, $|E_{\alpha}|$. The {\em empty} graph, the graph with no nodes and no links, will be depicted by the symbol $\{\}$.
The intersection graph ${\cal G}_{\cap}$ is defined as
\begin{equation}
{\cal G}_{\cap}=\bigcap_{\alpha\leq \mu}{\cal G}_\alpha,
\end{equation}
where the intersection symbol means
\begin{equation}
\bigcap_{\alpha\leq \mu}{\cal G}_\alpha\equiv {\cal G}(V_{\cap}, \bigcap_{\alpha\leq \mu} E_{\alpha}). 
\label{eq:bigcap}
\end{equation}
Here $V_{\cap}$ is the set of nodes which are at the endpoint of at least to one link in $\bigcap_{\alpha\leq \mu} E_{\alpha}$. 
Nodes that become isolated after the intersection operation are not considered for any of the computations involving ${\cal G}_\cap$. 
Note that the more levels the multiplex has, the more probable is that $|V|>|V_\cap|$.
One can of course intersect only specific layers of the multiplex. For the intersection of layers $\alpha_1, . . .,\alpha_k$ we write for the intersection graph
\[
{\cal G}^{\alpha_1, . . .,\alpha_k}_{\cap}=\bigcap_{\alpha_1, . . .,\alpha_k}{\cal G}_{\alpha_k} .
\]
Links in a given intersection graph are referred to as {\em multi-links} \cite{Bianconi:2013b}. In ${\cal G}^{\alpha_1, . . .,\alpha_k}_{\cap}$, 
two nodes are linked if they are linked in layers $\alpha_1, . . .,\alpha_k$. Links in ${\cal G}_{\cap}$ depict pairs of nodes which 
are connected through all the possible relations that define the multiplex --see figure \ref{fig:Schema}(a) and \ref{fig:Schema}(b).

\subsection{The ${\mathbf{G}}_K$-core and its community structure \label{sect2B}}
\subsubsection{The ${\mathbf{G}}_K$-core}
In the following we work with an intersection graph with layers that are considered relevant, for which we write ${\cal G}_{\cap}$. 
We then compute its {\em generalised} $K$-core, ${\mathbf{G}_K}$-core, which is defined as the maximal induced subgraph 
for which each node has either a degree equal or larger than $K$, {\em or} it connects two nodes whose degree is equal or larger than $K$. 
Recall that, as for the $\mathbf{K}$-core, the connectivity requirements must be satisfied inside the subgraph, so that a recursive algorithm must be used. The algorithm may work as follows: Starting with graph ${\cal G}$ we remove all  nodes $v_i\in {\cal G}$ satisfying that: (1) its degree is lower than $K$ {\em and} (2)  at most one of its nearest neighbours has degree equal or higher than $K$.  
We iteratively apply this operation over ${\cal G}$ until no nodes can be pruned, either because the derived subgraph is empty, or because all nodes which survived the iterative pruning mechanism cannot be removed following the above instructions. The graph obtained after this process is the {\em generalised} $K$-core subgraph, referred to as $\mathbf{G}_K$-core.
The inclusion of the connectors in the definition of the $\mathbf{G}_K$-core makes it a richer topological object. It has been shown that $\mathbf{G}_K$ is better suited for  the identification of the {\em elite}  in a social system than the standard $\mathbf{K}-core$ \cite{Corominas-Murtra:2014}.  

\subsubsection{The ${\mathbf{M}}$-core and the community structure in the core}

The $\mathbf{G}_K$-core can have internal structure itself  around core communities. We assume that {\em core communities} are 
formed by regions of the core which are highly clustered. 
The identification of highly clustered regions is performed by means of the $\mathbf{M}$-core \cite{Boguna:2013}. 
Given a graph ${\cal G}$, the $\mathbf{M}$-core of this graph, $\mathbf{M}({\cal G})$, is defined as the maximal induced 
subgraph of ${\cal G}$, in which {\em each link participates in at least} $M$ {\em triangles}. The $\mathbf{M}$-core highlights 
the role of triadic-closure within social dynamics, a process that seems to be a major driving force in social network formation 
\cite{Szell:2010b, Rapoport:1953, Granovetter:1973,  Davidsen:2002, Klimek:2013}.
In our case we will use it to identify the clustered regions of ${\mathbf{G}_K}({\cal G}_{\cap})$, which we denote by $\mathbf{MG}_K({\cal G}_{\cap})$.
Larger and lower values of $M$ will identify more or less clustered regions in the core, respectively. The different connected components of $\mathbf{MG}_K({\cal G}_{\cap})$ are the {\em core communities}. 

Finally, we point out that the identified  communities will in general not contain all the links associated  with the core; also some nodes may be removed in the process.  Formally this means that $\mathbf{G}_K({\cal G}_\cap)\setminus\mathbf{MG}_K({\cal G}_{\cap})\neq \{ \}$. This property will be relevant for the computation of the {\em Weak} core. 

\subsection{The Weak core and the Minimal Weak core \label{sect2C}}

%
The {\em Weak core} is the subgraph of the core in which all nodes and links participate in a path that goes from one core community to another, without crossing any of such communities. The weak core, thus, ensures the cohesiveness of the core of the network, acting as a gluing structure between core communities. 

We put the above informal statement in a more formal way, assuming the definitions of core and core community based upon the $\mathbf{G}_K$-core and $\mathbf{M}$-core, respectively. Let us assume that the core defined by ${\mathbf{G}_K}({\cal G}_{\cap})$ contains a single connected component 
and that the $\mathbf{MG}_K({\cal G}_{\cap})$ identifies several core communities  $C_1,C_2, . . .,C_m$ --which are, as we said above, the connected components of the $\mathbf{MG}_K$-core. 
The {\em weak core} of a multiplex graph, $\mathbf{W}_K({\cal G}_{\cap})$, is formed by  all links and nodes of $\mathbf{G}_K({\cal G}_{\cap})$ that participate in a path that starts at some node $v_k\in C_i$ and ends 
at some $v_{\ell}\in C_j$, where  $C_i$ and $C_j$ are different components of  ${\mathbf{MG}_K}({\cal G}_{\cap})$, 
with the constraint that all nodes in the paths but $v_k$ and $v_{\ell}$, if any, must belong to $\mathbf{G}_K({\cal G}_\cap)\setminus\mathbf{MG}_K({\cal G}_{\cap})$ --see figure 1b,c. The {\em Weak core} of a multiplex network is thus the region of the core of the intersection network 
which ensures the cohesiveness of the core. By definition the weak core itself is a weakly clustered region of the core, 
and its nodes may be among the least connected nodes of the core. In figure \ref{fig:Schema}(b) and \ref{fig:Schema}(c) we schematically show how such subgraphs can be derived.

We additionally define the {\em minimal weak core}, $\tilde{\mathbf{W}}_K$, as those links and nodes participating in all {\em minimal paths} from one component to an 
other in $\mathbf{MG}_K({\cal G}_{\cap})$. If there are two (or more) paths of $\mathbf{W}_K$ that connect $v_k\in C_i$ and $v_j\in C_x$, 
where $x\neq i$, we take the shortest. In case two or more paths connect such two nodes have the minimal length, we choose one at random. 
Note that by construction, if $\mathbf{W}_K\neq \{\}$, then:
\begin{eqnarray}
&&\mathbf{W}_K({\cal G}_{\cap})\cap \mathbf{MG}_K({\cal G}_{\cap})\neq \{\}\;{\rm and}\nonumber\\
&&\tilde{\mathbf{W}}_K({\cal G}_{\cap})\cap \mathbf{MG}_K({\cal G}_{\cap})\neq \{\}.\nonumber
\end{eqnarray}
The concept of the weak core is not tied to a particular definition of the {\em core} or a {\em core community}. 
One can define the core of a network in any suitable way (for example using the $K$-{core}). If it is possible to identify more than one community 
inside this core (using any method of community detection) the weak core is the region (links and nodes) that glues the communities. 
The reason by we suggest the combination of  the $\mathbf{G}_K$-{core} and the $\mathbf{M}_K$-{core} is that the first has been shown 
to perform better in  identifying relevant levels of core organisation (especially in social systems) than the classical $\mathbf{K}$-core,  
and because the $\mathbf{M}$-{core} captures clustering. It may happen that the $\mathbf{W}_K$-{core} is composed of a set of  
links that connect different components of the $\mathbf{MG}_K({\cal G}_{\cap})$, thereby indicating that all nodes in $\mathbf{G}_K({\cal G}_{\cap})$ 
are in $\mathbf{MG}_K({\cal G}_{\cap})$, and that the $\mathbf{M}$-{core} extraction only removed a few links. Finally, we say the Weak core is empty if the application  of the $\mathbf{M}$-{core} does not identify the communities within the $\mathbf{G}_K$-{core}.

\subsection{Identifying relevant levels of core organisation \label{sect2D}}
\begin{figure}
\begin{center}
\includegraphics[width=8.2cm]{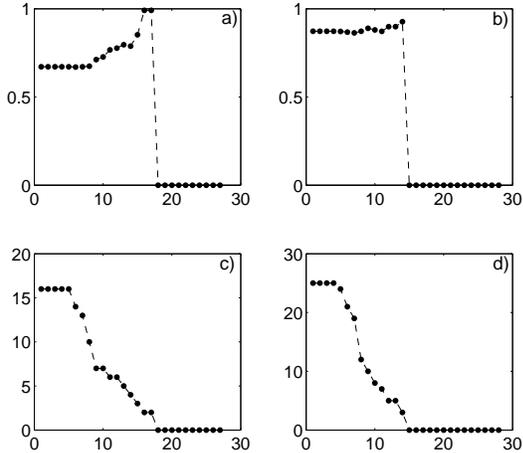}
\caption{Evolution of the relevance parameter $h(\mathbf{MG}_K)$ as a function of $K$ for the two periods under study; $796-856$ (a) and $1140-1200$ (b). In both periods we observe a remarkably constant behaviour with a slightly increasing trend followed by an abrupt decay. The larger is $h(\mathbf{MG}_K)$, the more relevant are the weak structures keeping the core connected. In (c) and (d) we plot the raw number of core communities of the cores of the two periods under study against $K$. We observe that such number decreases over time, although the increase on $h(\mathbf{MG}_K)$ tells us that the breaking is more and more uniform as long as $K$ increases, in terms of community sizes. Finally, the abrupt decay in $h(\mathbf{MG}_K)$ coincides with the fact that only one core community is identified, which occurs at deep levels of the core organisation. The $K$ level in which the $\mathbf{W}_K$-core is computed is the one displaying the maximum $h(\mathbf{MG}_K)$. In the first period, (a), the $\mathbf{G}_K$-core is already broken at the levels showing the maximum 
normalised entropy ($K=14,15$), we thus choose the largest $h(\mathbf{MG}_K)$ by which  the $\mathbf{G}_K$-core is not broken, $K=13$. }
\label{fig:h(GMK)}
\end{center}
\end{figure}

Which value of $K$ should be used to compute the $\mathbf{G}_K$-{core} such that the {\em weak core} reveals significant 
topological information of the organisation of the multiplex? Informally speaking, if the  $\mathbf{MG}_K$-{core} identifies a very large community and a set of other small commmunities, the role of the weak core will be less relevant than in 
the case where the communities, even eventually lower in number, have comparable size. 
The more uniform the size of the core communities, the more {\em relevant} are the level(s) for the core organisation. 

To identify such level(s), we compute the $\mathbf{MG}_K({\cal G}_{\cap})$ for all values of $K$ by which $\mathbf{MG}_K({\cal G}_{\cap})\neq \{\}$. 
For each of this levels, we proceed as follows:
Let $C_1, C_2,. . . ,C_m$ be the $m$ core communities of the $\mathbf{G}_K$-core, glued in this latter subgraph by means of the Weak core $\mathbf{W}_K({\cal G}_\cap)$.  
Let $|C_i|$ be the number of nodes of the component $M_i$ and let us define the probability that 
a randomly chosen node from $\mathbf{MG}_K({\cal G}_{\cap})$  is in the component $C_j$
\[
p(C_j)=\frac{|C_j|}{\sum_{i\leq m}|C_i|}. 
\]
We then compute the corresponding Shannon entropy 
\begin{equation}
H(\mathbf{MG}_K)=-\sum_{i\leq m}p(C_i)\log p(C_i).
\label{eq:H}
\end{equation}
The more uniform is the size distribution of the core communities, the larger will the entropy be. This enables us to compare different core community structures with the same number of components but with different community distribution sizes. For example, one can compare the situations where the $\mathbf{W}_K$-core glues 
two components of sizes $10$ and $100$, or $50$ and $50$. The role of the Weak core will be much more relevant within the core organisation in the second case than in the first one, and this is identified by the above entropy. 
To correct for size effects, we use the normalised Shannon entropy
\begin{equation}
h(\mathbf{MG}_K)=\frac{H(\mathbf{MG}_K)}{\log m}.
\label{eq:h}
\end{equation}
The most relevant level of core organisation, $K^\star$, if there exists any, will be located at the level $K$ for which
\begin{equation}
h(\mathbf{MG}_K)=\max_K\{h(\mathbf{MG}_K)\}.
\label{eq:hstar}
\end{equation}
If such a level exists, this will define the optimal value of $K$ with which the weak core will be computed.

Concerning the choice of  $M$, in the computation of the $\mathbf{MG}_K$-core, we use the following observation: 
If a given core does not break at low values of $M$, this means that the core is highly clustered and highly cohesive. 
In terms of the core organisation, the role of the community structure (if any)
will be less significant. We therefore choose $M$ as the minimum value that breaks the $\mathbf{G}_K$-core. 
Generally we will consider $M>1$, since values of $M=1$ can only isolate regions with low clustering and can not capture 
the idea of cohesive community. One can use other levels of $M$ to gain a better insight in the core structure of the graph.
  
\section{Results \label{sect3}}

We demonstrate the feasibility and quality of identifying the `connector regions' within the core of multiplex social systems 
with data from a social multiplex network of social interactions occurring in the virtual society of the  {\em Pardus} computer game.
The multiplex network is composed of cooperative interactions {\em friendship} ($F$),  {\em communication} ($C$) and {\em trade} ($T$). 
Our social system is therefore given by the MPN  ${\cal M}(t)={\cal M}(V,E_F\times E_C\times E_T,t)$, where 
$E_F, E_C$ and $E_T$ are the sets of links defining a friendship relation, a communication event, or a commercial relation, respectively. 
Our analysis is performed on the three networks  ${\cal G}_F, {\cal G}_C$ and ${\cal G}_T$  obtained from the most active players in 
two time windows of sixty days in length, $t_1=796$-$856$ and $t_2=1140$-$1200$. The time units here are days since beginning of the game. 
A link between two players in layer ${\cal G}_F$ exists if at least one player recognises the other as a `friend' within a time window.
A link between two players in layer ${\cal G}_C$ exists if at least one player has sent a message to the other, and 
a link between two players in ${\cal G}_T$ means that there has taken place at least one commercial transaction between the 
players in the time window. The set of players that defines the set $V$ of the MPN obtained from the period 796-856 contains 
2422 players,  and 2059 players for the period 1140-1200. 
Inactive players are removed from the MPN which leaves us with about $2000-2500$ players. Following equation (\ref{eq:bigcap}) and with these 
players we construct,  
\[
{\cal G}_{\cap} = {\cal G}_F\bigcap{\cal G}_C\bigcap{\cal G}_T.
\]
We drop the time label $T$ indicating the time window. All results are presented for ${\cal G}_{\cap}$. Single layer analysis or 
even intersections of two layers show much more noisy and unclear trends. ${\cal G}_{\cap}$ allows us to use the multiplex structure to 
reduce noise.

\subsection{Topological indicators \label{sect3A}}

In figure 2 we show the normalised entropy $h(\mathbf{MG}_K)$, equation  (\ref{eq:hstar}), as a function of  $K$ for both time windows in (a) and (b), respectively. 
The value of $h(\mathbf{MG}_K)$ remains almost constant with a slight increase before it abruptly jumps  to zero. 
This  constant plateau --see figure 2(a) and 2(b)-- is observed regardless 
if the number communities in the core of the network --see figure 2(c) and 2(d). It is true even the number of communities has significant variations --see figure 2(c) and 2(d). The number of communities shows  a decreasing trend until only one community is identified, provoking the collapse of $h(\mathbf{MG}_K)$ to zero. Note that the collapse occurs just after the value of $K$ at  which the communities of the core have comparable size. 
If only a single layer of the multiplex system is used, the situation is less well pronounced than the case shown in figure 2 (a) and (b).
Relevant levels identified using the procedure described in section \ref{sect2D} are found for $K=14$ for the first period, $796$-$856$ days, and $K=13$ for the second, $1140$-$1200$ days. 
Although for the first period $h(\mathbf{MG}_K)$ is higher for $K=15$ and $16$, at these stages the $\mathbf{G}_K$-core is already 
broken into two components, whereas at $K=14$ it still contains one single component, as required by the proposed method.

To compute the $\mathbf{MG}_K$-core we set $M=2$. The $\mathbf{MG}_K$-core detects three highly clustered communities 
of comparable size in both periods, containing $68\%$ and $61\%$ of all nodes of the $\mathbf{G}_K$-core in the first and second time period, respectively. 
These communities show a high clustering coefficient $c \sim 0.6-0.7$ (clustering of the $\mathbf{G}_K$-cores is $\sim 0.5$), 
and an average degree of around $\langle k \rangle \sim 7$,  which is similar to the average degree of the ${\mathbf{G}_K}$-cores in both periods.
The relative sizes of the identified weak cores in relation to the respective $\mathbf{G}_K$-cores are $0.27$ and $0.28$ for the first and second periods, respectively. 
The $\mathbf{W}_K$-core is formed by a weakly connected region exhibiting less than $1/2$ of the average degree of the $\mathbf{G}_K$-core,  
$\langle k\rangle_{\mathbf{W}}=3.0$ and $\langle k\rangle_{\mathbf{GK}}=7.0$, and $\langle k\rangle_{\mathbf{W}}=2.9$ and $\langle k\rangle_{\mathbf{GK}}=6.8$ 
in two periods, respectively. As expected the clustering is almost vanishing around $c \sim0.07$ in both periods. 

The most surprising topological property of the observed weak cores is that, in spite their low connectivity and their role as 
connector regions, they define a single connected component in both time periods. This reveals that the $\mathbf{W}_K$-core plays an important 
functional role in the underlying the organisation of the network.  We find that in both time periods $\tilde{\mathbf{W}}_K\approx \mathbf{W}_K$. 
This means that the raw $\mathbf{W}_K$-core is quite optimal in the sense that a few redundant paths connecting the communities of the $\mathbf{G}_K$-core are identified.  This confirms the property of the identified weak core as a true minimal gluing region that keeps the cohesiveness of the core of the multiplex network. 

\subsection{Social Performance Indicators \label{sect3B}}
Social indicators and social performance measures of those players that populate the weak core show interesting and unexpected results. 
These indicators are: {\em Experience} is a numerical indicator accounting for the experience of the player, related to battles in which the player has participated, 
or the number monsters he/she killed.
{\em Activity} is a numerical indicator related to the number and complexity of actions performed by the player.
{\em Age} is the number of days after the player joined the game.
Finally, {\em wealth} is a numerical indicator accounting for the wealth of the player within the game at any point in time. 
Wealth accounts for money and the cumulative value of a payers's equipment within the game.
We list the average experience level, activity level, age and wealth of those nodes in table 1.

The most salient observation is that for almost all indicators in the two periods under study, those nodes that compose the weak core 
have the highest social scores when compared to nodes composing the core,  its clustered communities, or the average player. 
Even the communities of the core are defined by a strong connectivity pattern, which does not guarantee the best performance in social indicators. 
This tells us that being located  between different core communities leads to superior social performance. 
We find one exception where the wealth in the second period is higher for the core communities.  
In addition, one finds that the age of the players populating the weak core is sensibly larger than the average wealth of the core 
and than its communities defined by the $\mathbf{MG}_K$-core. In table 1 we collect the results, highlighting the best scores.
\begin{table}[!ht]
\caption{Table with the social indicators of the different subgraphs of ${\cal G}_{\cap}$ corresponding to periods $t_1=796$-$856$ and $t_2=1140$-$1200$}\footnote{For the first period, $\tilde{\mathbf{W}}_K=\mathbf{W}_K$.}
Ê\begin{tabular}{ccccc}
\hline
{\rm Subgraph}&$\langle {\rm Experience}\rangle$&$\langle {\rm Activity}\rangle$&$\langle {\rm Wealth}\rangle$&$\langle {\rm Age}\rangle$\\
\hline
\hline
$t_1$&&&\\
\hline
$\mathbf{G}_K$& $4.9\times 10^5$ & $3.63\times 10^6$  & $5\times 10^7$ & $677$ \\
$\mathbf{MG}_K$ & $4.77\times 10^5$ & $3.62\times 10^6$ & $4.88\times 10^7$   &$668$\\
$\mathbf{W}_K$ & $\mathbf{6.01\times 10^5}$ & $\mathbf{4.11\times 10^6}$ & $\mathbf{5.18\times 10^7}$ & 732  \\
$\tilde{\mathbf{W}}_K$ & $\mathbf{6.01\times 10^5}$ & $\mathbf{4.11\times 10^6}$ & $\mathbf{5.18\times 10^7}$  & 732 \\
\hline
$t_2$&&&\\
\hline
$\mathbf{G}_K$ & $7.72\times 10^5$ & $5.69\times 10^6$  & $9.84\times 10^7$ & 1020\\
$\mathbf{MG}_K$ & $8.58\times 10^5$ & $6.14\times 10^6$ & $\mathbf{1.13\times 10^8}$ &1060 \\
$\mathbf{W}_K$& $1.02\times 10^6$ & $6.3\times 10^6$  & $9.85\times 10^7$  &1030 \\
$\tilde{\mathbf{W}}_K$ & $\mathbf{1.03\times 10^6}$ & $\mathbf{6.38\times 10^6}$ & $1.04\times 10^8$ &1070\\
\hline
\hline
Ê\end{tabular}
\end{table}
We finally note that in the second period the $\mathbf{MG}_K$-core is already broken into three components 
$C_1,C_2,C_3$ for $M=1$. Remarkably, the weak core is formed only by two links, that connect 
$C_1$ with $C_2$, and $C_1$ with $C_3$. This identifies what one could call {\em supercritical links} at the core of the multiplex society.

\section{Discussion \label{sect4}}

In this paper we described a new type of subgraph, the weak core, which belongs to the family of core subgraphs. The latter include the 
clique subgraphs \cite{Harary:1957}, the {\em Rich club} \cite{Colizza:2006}, the standard $\mathbf{K}$-core \cite{Seidman:1983, Bollobas:1984, Dorogovtsev:2006}, 
and the generalised $\mathbf{K}$-core \cite{Corominas-Murtra:2014} as well as other approaches \cite{Corominas-Murtra:2007, Corominas-Murtra:2008}. 
The interest of this weak core arises since it captures a property that is essential for the identification of elite structure in social systems: 
The ability of the high social performers to maintain ties to the various core communities that organise the whole topology of the system from its core. 
The core of the multiplex network, defined as the generalised $K$-core of the intersection network from all layers in the MPN provides a 
rich structure in which one can identify core communities. In our case, we identified the community structure of the core of the MPN through the application of the $\mathbf{M}$-core. In doing so, we consider that core communities are defined by those regions of the core which depict a highly clustered structure. In a totally opposite way, the weak core is comprised of regions of the core that are neither highly connected nor well clustered. 
This region's primary role is to keep the cohesiveness of the core.

The weak core identifies those individuals performing best in the virtual society. In previous studies, it has been shown that there is a direct relation 
between the degree of the player and its performance \cite{Fuchs:2014}. However, our findings indicate that nodes that are high social performers, well 
connected and part of a core group,  need ties to other communities in the core. The weak core suggests a deeper structure of elites in social systems, 
and includes what seems to be a crucial for elite members: the ability to maintain ties beyond the community they belong to. 
Moreover, some members of the weak core may not belong to any core community and their role within the core organisation is purely devoted to keep the cohesive nature of it. This role as topological hinge between core communities may lead this particular class of players to an increase of their social performance.

The presented methodology is not tied to the particular definitions of the core or core community. Further works should stress 
the functional role of these weakly connected regions at the core of multiplex systems. In addition, the notion of weak core could 
be applied to other fields where this type of brokerage structure may play an important role in organising networks, such as in 
neurology or biological networks.

\acknowledgments
This work was supported by the Austrian Science Fund FWF under KPP23378FW, the EU LASAGNE project, no. 318132 and the EU MULTIPLEX project, no. 318132.

\end{document}